\documentstyle[twoside,twocolumn,prd,aps]{revtex}
\begin{document}
\draft
\title{{Reply to `Comment on `Gravitating Magnetic Monopole in the Global Monopole 
Spacetime' '}}
\author{E. R. Bezerra de Mello \thanks{E-mail: emello@fisica.ufpb.br}}
\address{{Departamento de F\'{\i}sica-CCEN
Universidade Federal da Para\'{\i}ba
58.059-970, J. Pessoa, PB
C. Postal 5.008, Brazil}\\
\parbox{14 cm}{\medskip\rm\indent
In this Reply I present some arguments in favor of the stability of the topological 
defect composed by global and magnetic monopoles.}}
\maketitle

In a recent paper A. Ach\'ucarro and J. Urrestilla \cite{AU} have pointed the 
stability problem for composite, global and magnetic monopoles defect, presented 
in a previous publication \cite{JUE}. In the former it is claimed that this 
topological defect is not classically stable against axially symmetric angular 
deformation on the boson field associated with the global sector \footnote{The 
stability problem of the global monopole against angular perturbation has been 
observed  by A. S. Goldhaber \cite{G}, and also by the authors in \cite{AU1}.}.
By this deformation an extra tension is created at the north-pole which drags the 
core of the global monopole upwards {\it with no cost of energy}. Once the cores of 
both monopoles are separated, they are repelled by an induced self-interaction 
consequence of the distortion on the magnetic fields due to the solid angle deficit
$\delta\Omega=4\pi\Delta$, being $\Delta$ the parameter associated with the energy
scale where the gauge symmetry is spontaneously broken. However, in the composite 
defect system this analysis must take into account the induced self-energy more 
carefully. In order to infer the behavior of this self-energy I briefly comment 
previous analysis for the ideal case. In \cite{EC} it was 
calculated the formal expression to the electrostatic self-energy associated with a 
test charged particle in the pointlike global monopole spacetime. It was observed 
that it is positive, and by numerical evaluation can be seen that the induced
electrostatic strength increases for larger values of the parameter $\Delta$.
\footnote{Although in \cite{AU} the authors said that the 
effect of the core has been considered, this is not true. In \cite{EC} the metric 
tensor considered was associated with the global monopole with no internal 
structure.} Considering the effect of the core of the 
global monopole, the calculation of the self-energy becomes much more complicated.
Although the exact expression for this self-energy becomes impossible to be obtained, 
we can infer its behavior as following: 
at the global monopole center, $r=0$, the metric tensor is
Minkowski. Its component $g_{11}$ changes from $1$ to $1/(1-\Delta)$ for distance
$r\geq r_{gm}$, increasing the value of an effective solid angle deficit. On the
other hand, the distortion on the electric field decreases for larger separation. So
for the case where increasing the distance the augment of the solid angle deficit
becomes dominant, the distortion increases with the separation up to $r_{gm}$;
consequently we can infer that this self-energy goes to zero for very small 
separation between the charged particle and the core of the monopole, peaks around 
the distance of the order of the monopole's size, $r_{gm}$, and decreases 
approximately with $1/r$ for larger value of separation. Assuming that this effect 
is presented to magnetic charged particle as the authors also admit, to separate the 
cores of the global and magnetic monopoles needs some amount of energy to overcome 
this barrier \footnote{An estimation for the peak of the magnetostatic self-energy 
can be provided considering the expression for self-energy given in \cite{EC} and 
that the monopole's size is of order $\eta^{-1}\lambda^{-1/2}$ \cite{BV}.}. So the 
question about the classical stability condition of the composite defect against 
axially symmetric perturbation must take into account this fact.

Another point raised by  Ach\'ucarro and Urrestilla is about the profile of 
the Higgs field. In fact for the model analysed in \cite{JUE} there are 
three mass parameters to scale the radial distance. In our numerical analysis we 
adopted the mass of the boson vector: $x=e\eta r$. This choice of mass
scale seemed convenient for us to analyse the behavior of the fields varying the 
self-coupling $\lambda$, keeping the electric coupling $e$ fixed.

\noindent
{\bf Acknowledgment}\\
\noindent
This work was partially supported by CNPq.

\end{document}